\documentclass[pdflatex,sn-mathphys-num]{sn-jnl}

\usepackage{graphicx}%
\usepackage{multirow}%
\usepackage{amsmath,amssymb,amsfonts}%
\usepackage{amsthm}%
\usepackage{mathrsfs}%
\usepackage[title]{appendix}%
\usepackage{xcolor}%
\usepackage{textcomp}%
\usepackage{manyfoot}%
\usepackage{booktabs}%
\usepackage{algorithm}%
\usepackage{algorithmicx}%
\usepackage{algpseudocode}%
\usepackage{listings}%

\theoremstyle{thmstyleone}%
\theoremstyle{thmstyletwo}%

\theoremstyle{thmstylethree}%

\begin{document}

\title[Article Title]{Dynamical System Analysis of Chameleon Mechanism in Brans-Dicke Scalar-Tensor Model}


\author*[1,2]{\fnm{Azwar} \sur{Sutiono}}\email{azwarsu@gmail.com}

\author[1,3]{\fnm{Agus} \sur{Suroso}}\email{suroso@itb.ac.id}
\equalcont{These authors contributed equally to this work.}

\author[1,3]{\fnm{Freddy} \sur{Permana Zen}}\email{fpzen@itb.ac.id}
\equalcont{These authors contributed equally to this work.}

\affil*[1]{\orgdiv{Theoretical Physics Laboratory, THEPI Division Indonesia}, \orgname{Institut Teknologi Bandung}, \orgaddress{\street{Jl. Ganesha 10}, \city{Bandung}, \postcode{40132}, \country{Indonesia}}}

\affil[2]{\orgdiv{Department of Physics}, \orgname{Universitas Hasanuddin}, \orgaddress{\street{Jl. Perintis Kemerdekaan KM. 10}, \city{Makassar}, \postcode{90245}, \country{Indonesia}}}

\affil[3]{\orgdiv{Indonesia Center for Theoretical and Mathematical Physics (ICTMP)}, \orgname{Institut Teknologi Bandung}, \orgaddress{\street{Jl. Ganesha 10}, \city{Bandung}, \postcode{40132}, \country{Indonesia}}}


\abstract{We investigated the stability of the chameleon screening mechanism in the Brans-Dicke scalar-tensor model. We define a constraint on the Brans-Dicke parameter $\omega_{BD}^*$ identifying two stability groups, $\omega_{BD}>\omega_{BD}^*$ and $0<\omega_{BD}<\omega_{BD}^*$. The first group achieves stability with both appropriate eigenvalues and a density profile consistent with dark energy dominance. The second exhibits eigenvalue stability but contradicts conditions for a stable universe. We explore the impact of variations in the scalar field potential and matter coupling by analyzing different parameter sets. Each unique set of parameters results in a distinct $\omega_{BD}^*$. Dynamic analysis reveals that stability is achieved when the scalar field dominates, highlighting the importance of the kinetic and potential terms while minimizing the influence of matter density. In high matter density regions, the scalar field's negligible presence aligns with standard gravitational behavior, whereas in low matter density regions, the scalar field grows exponentially, driving dark energy and cosmic acceleration.}




\maketitle

\section{Introduction}\label{sec1}
Modified gravity refers to theoretical modifications or alternative theories to Einstein's general theory of relativity, which describes the force of gravity as the curvature of spacetime caused by mass and energy \cite{cop}. While general relativity has been extremely successful in explaining various gravitational phenomena, there are some observations on cosmological and galactic scales that it doesn't fully account for. Modified gravity theories attempt to address these issues by introducing new concepts, fields, or modifications to the gravitational force. The field of modified gravity was still an active area of research, with scientists exploring various theoretical frameworks and observational tests. One of the most significant observations in cosmology is that the expansion of the universe is accelerating. This discovery, based on observations of distant supernovae, led modified gravity to the proposal of dark energy as a hypothetical form of energy that permeates space and counteracts the attractive force of gravity \cite{clif}\cite{riess}\cite{adam}\cite{perl}\cite{kom}\cite{hins}. Some examples of modified gravity theories include Scalar-Tensor Theories, $f(R)$ Gravity, Extra Dimensions and Braneworlds, and the Tensor-Vector-Scalar (TeVeS) Theory \cite{amen}\cite{gom}\cite{cald}\cite{soc}\cite{sde}\cite{koy}\cite{zha}\cite{oli}. 

Screening mechanisms come into play within these modified gravity theories. They are mechanisms that suppress or screen the effects of the modified gravity in environments where standard gravity predictions are well-established. This screening allows the theory to be consistent with local tests of gravity (such as those conducted in our solar system) while still producing noticeable effects on cosmic scales. Common types of screening mechanisms include the chameleon mechanism, Vainshtein mechanism, and symmetron mechanism \cite{kim}\cite{dim}\cite{hin}\cite{oliv}. These mechanisms allow modified gravity theories to evade detection in regions where gravitational effects are well-understood, ensuring agreement with experimental tests conducted in our immediate cosmic neighborhood.  The background of screening mechanisms is rooted in the quest to reconcile cosmological observations, particularly the accelerated expansion of the universe, with modified gravity theories that can explain these phenomena without relying on the introduction of dark energy. These mechanisms enable consistency with local tests of gravity while allowing for deviations at cosmic scales \cite{kho}\cite{khou}\cite{brax}\cite{joy}\cite{water}.

The Chameleon mechanism, embedded in the Brans-Dicke scalar-tensor model \cite{bran}\cite{bis}\cite{saai}\cite{lu}, adeptly conceals gravitational modifications in regions of high density. This mechanism marks a progression from the conventional chameleon mechanism, intricately coupling the scalar field with matter \cite{naka}\cite{sut}. As a consequence, the scalar field assumes a variable mass contingent upon the density of matter in the relevant region. In regions characterized by extremely low matter density, the scalar field within the model reverts to its role as dark energy, contributing to the universe's expansion acceleration \cite{sut}. 

Equally crucial, the pursuit of screening mechanism stability in modified gravity originates from the need to understand the behavior and viability of these theories in different environments, particularly on cosmological and astrophysical scales. The indispensability of stability becomes evident in ensuring the viability of modified gravity theories. An unstable mechanism could manifest observable deviations from general relativity in regions subject to experimental and observational constraints. Stability ensures that the modified gravity theory remains consistent with empirical evidence and can provide a viable alternative explanation for gravitational phenomena. The exploration of screening mechanism stability serves as a crucial means to anticipate and understand how these theories behave in diverse cosmic environments \cite{roy}\cite{nand}\cite{pal}. It offers insights into the interplay between gravity and matter at different energy scales and provides a testing ground for the compatibility of modified gravity theories with the principles of general relativity. Analogous to conventional dark energy models, the Chameleon mechanism must satisfy stringent stability conditions to establish itself as a trustworthy dark energy model. Hence, a thorough analysis of the dynamic system is conducted to assess the stability of the mechanism.

The structure of the paper is outlined as follows. Section \ref{sec2} give review of the thin-shell solution of the Brans-Dicke chameleon.  Section \ref{sec3} delves into the Brans-Dicke chameleon model, presenting the system of equations formulated as an autonomous system. Moving to Section \ref{sec4}, a comprehensive system dynamics analysis is conducted to assess the stability of the mechanism. Section \ref{sec5} elucidates the correlation between the Brans-Dicke chameleon and the universe's expansion acceleration. The final section encapsulates a summary and draws conclusions based on the findings of the study.

\section{Thin-shell Solution of Brans-Dicke Chameleon}\label{sec2}
In our previous work, we investigate the thin-shell solution of the Brans-Dicke Chameleon mechanism \cite{sut}. The Brans-Dicke Chameleon action is given as follows:
\begin{equation}
S=\int d^4x\sqrt{-g}\left[\phi \mathcal{R}-\frac{\omega_{BD}}{\phi}(\partial \phi)^2-2V(\phi) \right] + 2S_m\left(A^2(\phi)g_{\mu\nu}, \psi\right),\label{eq1}
\end{equation}
where $(\partial\phi)^2=g^{\mu\nu}\partial_{\mu}\phi\partial_{\nu}\phi$, $V(\phi)$ is the scalar field self-interaction potential, $w_{BD}$ is the Brans-Dicke dimensionless coupling parameter, $A(\phi)$ is the coupling function to matter, $g_{\mu\nu}$ is the metric in Jordan Frame and $\tilde{g}_{\mu\nu}=A^2(\phi)g_{\mu\nu}$ is the metric in Einstein Frame. The potential $V(\phi)$ is assumed to be of the runaway form that monotonically decreasing and satisfies: $V,\,V_{,\phi}/V,\,V_{,\phi\phi}/V_{\phi}...\rightarrow \infty$ as $\phi\rightarrow \infty$, as well as $V,\,V_{,\phi}/V,\,V_{,\phi\phi}/V_{\phi}...\rightarrow \infty$ as $\phi\rightarrow 0$ \cite{khou}\cite{brax}. By appropriately selecting $V(\phi)$ and $A(\phi)$, it is feasible to create a scenario where the scalar propagates unrestrictedly and facilitates a fifth force in regions characterized by a low Newtonian potential. However, this force is deactivated in high-density areas, such as those found in the solar system. Variation of action (\ref{eq1}) with respect to $\phi$ give the equation of motion for $\phi$,
\begin{equation}
\nabla^2\phi=\frac{2}{3+2\omega_{BD}}\left[\phi V_{,\phi}-2V+\left(A_{,\phi}\phi-\frac{A}{2}\right)\rho\right],\label{eq2} 
\end{equation}
with $\rho$ is density in general. In order to understand how the chameleon force is suppressed in the presence of high density region, we solve for the field profile in the presence of a massive compact object. We consider a spherically symmetric object characterized by a constant radius $R$ and density $\rho_c$ for simplicity. Additionally, we envision that this object exists in the presence of a homogeneous background of density $\rho_b$. We choose the thin-shell case that refers to a region around a massive object where the chameleon field experiences a significant change in its behavior. This thin shell arises due to the interaction between the chameleon field and matter, particularly in regions of high density. Within the thin shell, which typically extends to a certain distance from the surface of the massive object (its distance is $R_c$ from the center of the massive object), the chameleon field transitions from its light, weakly interacting state to its heavier, strongly interacting state. This transition zone effectively shields the effects of the chameleon field from being observable at larger distances.\\

We find the approximate  exterior solution of the thin-shell case,
 \begin{equation}
\phi_{thin}(r)=-\eta \frac{e^{-m_b(r-R)}}{r}+\phi_b, \label{eq3}
\end{equation} 
with,
\begin{eqnarray}
    \eta&=&\frac{e^{-(R_c+R)\sqrt{\frac{Q\beta\rho_c}{M_{pl}}}}(M_{pl}-2\beta) }{2\beta}\left(\frac{e^{2R_c \sqrt{\frac{Q\beta\rho_c}{M_{pl}}}}M_{pl}R_c\left(-R_c+\sqrt{\frac{M_{pl}}{Q\beta\rho_c}}\right)\left( 1+R\sqrt{\frac{Q\beta\rho_c}{M_{pl}}}\right)}{-R_c\left(Q\beta\rho_c+M_{pl}\right)+M_{pl}\left(\sqrt{\frac{M_{pl}}{Q\beta\rho_c}}- \sqrt{\frac{Q\beta\rho_c}{M_{pl}}}\right)}\right.\nonumber \\
&&\left.+\frac{e^{2R \sqrt{\frac{Q\beta\rho_c}{M_{pl}}}}R_c\left(R+\sqrt{\frac{M_{pl}}{Q\beta\rho_c}}\right)\left(Q R_c\beta\rho_c+\sqrt{\frac{Q\beta\rho_c}{M_{pl}}}\right)}{\sqrt{\frac{\beta\rho_c}{M_{pl}}}\left(R_c\left(Q\beta\rho_c-M_{pl}\right)+M_{pl}\left( \sqrt{\frac{M_{pl}}{Q\beta\rho_c}}+ \sqrt{\frac{Q\beta\rho_c}{M_{pl}}}\right)\right)}\right),\label{eq4}
\end{eqnarray}
where $Q=\frac{2}{3+2\omega_{BD}}$, $M_{pl}$, $\beta$, $\phi_b$ and $m_b$  are Planck mass, coupling constant, a minimum of the effective potential in the background, and the mass of small fluctuations about $\phi_b$, respectively. Note that $\eta$ is a constant that depends on Planck mass, theory parameters, and the compact object radius and density.

The solution is subsequently implemented on earth by choosing the suitable constants. As observed, the chameleon field undergoes exponential growth once it surpasses the Earth's radius, eventually stabilizing at an infinite distance. In regions with high matter density, where the chameleon field has a large effective mass, the associated fifth force becomes short-ranged. The force acts effectively only over small distances, confined to the immediate vicinity of the massive object. The short-range nature of the fifth force leads to a screening effect. On the contrary, in low-density regions far from the Earth, where a lighter scalar field operates, a long-range force is mediated, highlighting the impact of modified gravity on the cosmological background.

\section{Autonomous System of Brans-Dicke Chameleon Action}\label{sec3}
Considering the variation of the Chameleon Brans-Dicke action with respect to both the metric $g_{\mu\nu}$ and the scalar field $\phi$ in the flat Friedmann–Lemaître–Robertson–Walker background, we arrive at the following Friedmann equation and dynamic equation:
\begin{eqnarray}
3H^2&=&-\frac{3H\dot{\phi}}{\phi}+\frac{\omega_{BD}}{2}\left(\frac{\dot{\phi}}{\phi}\right)^2-\frac{V}{\phi}+\frac{A\rho_m}{2},\label{eq5}\\
2\dot{H}+3H^2&=&-\frac{2H\dot{\phi}}{\phi}-\frac{\omega_{BD}}{2}\left(\frac{\dot{\phi}}{\phi}\right)^2-\frac{\ddot{\phi}}{\phi}-\frac{V}{\phi},\label{eq6}\\
\ddot{\phi}+3H\dot{\phi}&=&\frac{2}{3+2\omega_{BD}}\left[\phi V_{,\phi}-2V+\left(A_{,\phi}\phi-\frac{A}{2}\right)\rho_m\right].\label{eq7}
\end{eqnarray}
where $\rho_m$ represents the matter density. The matter fluid is considered as pressureless dust $(p_m = 0)$. In cosmology, pressureless dust is a good approximation for non-relativistic matter, such as cold dark matter and baryonic matter, at large scales. These components dominate the matter content of the universe in the current epoch, making pressureless dust a relevant model for studying the chameleon mechanism \cite{roy}. In the absence of a screening effect,  without any coupling between matter and the scalar field, the analysis reduces to the standard Brans-Dicke theory \cite{nand}. 
Combining equation (\ref{eq5}), (\ref{eq6}) and (\ref{eq7}), we get modified continuity equation as,
\begin{eqnarray}
    \dot{\rho}_m+3H\rho_m=-\frac{3}{4}\frac{A_{,\phi}\dot{\phi}\rho_m}{A}
\end{eqnarray}The effective equation of state of the system is $\omega_{eff}=-1-\frac{2}{3}\frac{\dot{H}}{H^2}$ \cite{amen}. In order to study the cosmological dynamics of the system, it is convenient to introduce the following dimensionless variables: 
\begin{equation}
 x=\frac{\dot{\phi}}{H\phi},
 ~~~~~y=\frac{1}{H}\sqrt{\frac{V}{3\phi}},
 ~~~~~z=\frac{1}{H}\sqrt{\frac{\rho_mA}{3\phi}}.\label{eq8}
\end{equation}
The equation (\ref{eq5}) can be simplified and written as a constraint,
\begin{equation}
\frac{\omega_{BD}}{6} x^2 -x - y^2 +z^2 = 1.\label{eq9}
\end{equation}
We also define the energy fraction of dark energy and matter:
\begin{equation}
    \Omega_{\phi}=\frac{\omega_{BD}}{6}x^2-x - y^2;\,\,\,\,\,\,\Omega_m=1-\Omega_{\phi}.\label{eq10}
\end{equation}
From equation (\ref{eq6}) and (\ref{eq7}), we get
\begin{eqnarray}
    \frac{\dot{H}}{H^2}&=&-\frac{x}{2}-\frac{\omega_{BD}x^2}{4}+3\left( Q-\frac{1}{2} \right)y^2+\frac{3Qz^2}{4}-\frac{Q}{2H^2}\left(V_{,\phi}+A_{,\phi}\rho_m\right)-\frac{3}{2},\label{eq11}\\
    \ddot{\phi}&=&-3H\dot{\phi}-6Qy^2H^2\phi-\frac{3Qz^2H^2\phi}{2}+Q\phi\left( V_{,\phi}+A_{,\phi}\rho\right).\label{eq12}
\end{eqnarray}

By choosing $V(\phi)=V_0\phi^{-n}$ and $A(\phi)=A_0\phi^{\alpha}$, with $n$ and $\alpha$ are positive dimensionless constant, then differentiating each dimensionless variable in equation (\ref{eq8}) with respect to the number of e-foldings $N=$ ln $a$, obtain the autonomous equations\\
\begin{eqnarray}
\frac{dx}{dN}&=& \frac{\omega_{BD}x^3}{4}-\frac{3x^2}{2}-\frac{3x}{2}-3\left(n+2\right)Qy^2+3\left( \alpha-\frac{1}{2}\right)Qz^2\nonumber\\
&&-3\left[\left(\frac{n}{2}+1\right)Q-\frac{1}{2} \right]xy^2+\frac{3}{2}\left( \alpha-\frac{1}{2}\right)Qxz^2,\label{eq13}\\
\frac{dy}{dN}&=&-3\left[\left(\frac{n}{2}+1\right)Q-\frac{1}{2} \right]y^3+\frac{3y}{2} +\frac{\omega_{BD}x^2y}{4}-\left(\frac{n}{2}+1 \right)xy\nonumber\\
&&+\frac{3}{2}\left( \alpha-\frac{1}{2}\right)Qyz^2,\label{eq14}\\
\frac{dz}{dN}&=&\frac{3}{2}\left(\alpha-\frac{1}{2} \right)Qz^3 +\frac{\omega_{BD}x^2z}{4}-3\left(\frac{n}{2}+1 \right)Qy^2z+\frac{3y^2z}{2}\nonumber\\
&&+\left(\frac{\alpha}{8}-1\right)xz,\label{eq15}
\end{eqnarray}
and effective equation of state,
\begin{eqnarray}
    \omega_{eff}&=&-1-\frac{2}{3}\left[-\frac{x}{2}-\frac{\omega_{BD}x^2}{4}+3\left( Q-\frac{1}{2} \right)y^2+\frac{3Qz^2}{4}\right. \nonumber \\
&&\ \left.+\frac{3Q}{2}\left(ny^2 -\alpha z^2 \right)-\frac{3}{2} \right],\label{eq16}
\end{eqnarray}
with $Q=\frac{2}{3+2\omega_{BD}}$ and $\frac{d}{dN}=\frac{1}{H}\frac{d}{dt}$. By setting $\frac{dx}{dN}=\frac{dy}{dN}=\frac{dz}{dN}=0$, fixed points can be obtained. Solving the above equations we find the fixed points that some explicitly depend on $\omega_{BD}$, $n$, and $\alpha$ as illustrated in Table \ref{tabA}. The dynamics of the autonomous system can be analyzed by finding the exact constant fixed points of the system and examining the stability around these points \cite{bah}.

\begin{table}[h
]
\caption{Fixed points of autonomous equation depend on $\omega_{BD}$, $n$, and $\alpha$.}\label{tabA}%
\begin{tabular}{@{}cccc@{}}
\toprule
Point & $x$  & $y$ & $z$\\
\midrule
P1    & $0$   & $0$  & $0$  \\
P2    & $\frac{12}{4n+\alpha}$   & $-\frac{A}{B}$  &  $-\frac{C}{D}$ \\
P3    & $\frac{12}{4n+\alpha}$   & $-\frac{A}{B}$  &  $\frac{C}{D}$  \\
P4    & $\frac{12}{4n+\alpha}$   & $\frac{A}{B}$ &  $-\frac{C}{D}$  \\
P5    & $\frac{12}{4n+\alpha}$   & $\frac{A}{B}$  &  $\frac{C}{D}$  \\
P6    & $\frac{6}{2n+1}$   & $-E$  & $0$  \\
P7    & $\frac{6}{2n+1}$    & $E$  & $0$  \\
P8    & $\frac{4-2\alpha}{4+\alpha+4\omega_{BD}}$   & 0  & $-F$  \\
P9    & $\frac{4-2\alpha}{4+\alpha+4\omega_{BD}}$   & 0  & $F$  \\
P10    & $\frac{2\left(n+2\right)}{1-n+2\omega_{BD}}$   & $-G$  & $0$  \\
P11    & $\frac{2\left(n+2\right)}{1-n+2\omega_{BD}}$   & $G$  & $0$  \\
P12    & $\frac{3-\sqrt{3}\sqrt{3+2\omega_{BD}}}{\omega_{BD}}$   & $0$  & $0$  \\
P13    & $\frac{3+\sqrt{3}\sqrt{3+2\omega_{BD}}}{\omega_{BD}}$   & $0$  & $0$  \\
\botrule
\end{tabular}
\footnotetext{$A=\sqrt{-24+8n-24\omega_{BD}-4\alpha-4n\alpha-\alpha^2}$}
\footnotetext{$B=\sqrt{16n^2+8n\alpha+\alpha^2}$}
\footnotetext{$C=i\sqrt{2\left(\alpha-2\right)\left(-6+14n+4n^2+2\alpha+n\alpha-12\omega_{BD}\right)}$}
\footnotetext{$D=\sqrt{-16n^2-8n\alpha+32n^2\alpha-\alpha^2+16n\alpha^2+2\alpha^3}$}
\footnotetext{$E=\frac{\sqrt{3\left(n-1\right)\left(2\omega_{BD}+3\right)+\frac{1}{2n+1}\left[36-18n\left(n+1\right)+\left(78-12n-12n^2+36\omega_{BD}\right)\omega_{BD}\right]}}{\sqrt{-3-3n+6n^2-6\omega_{BD}-12n\omega_{BD}}}$}
\footnotetext{$F=\frac{\sqrt{\frac{2-\alpha}{\alpha+4\omega_{BD}+4}\left(192+272\omega_{BD}+96\omega_{BD}^2+24\alpha+16\omega_{BD}\alpha-6\alpha^2-4\omega_{BD}\alpha^2\right)}}{{\sqrt{24\alpha^2+96\omega_{BD}\alpha+84\alpha-48\omega_{BD}-48}}}$}
\footnotetext{$G=\frac{\sqrt{3\left(n-1\right)\left(2\omega_{BD}+3\right)+\frac{n+2}{1-n+2\omega_{BD}}\left[12-6n\left(n+1\right)+\left(26-4n-4n^2+12\omega_{BD}\right)\omega_{BD}\right]}}{\sqrt{-3\left(n+1\right)+6n^2-6\left(2n+1\right)\omega_{BD}}}$}
\end{table}

\section{Stability Analysis of the Brans-Dicke Chameleon}\label{sec4}

The stability of a fixed point in a dynamical system is characterized by how the system behaves in the vicinity of that point. Specifically, stability analysis involves understanding the response of the system to small perturbations or deviations from the fixed point. In the context of autonomous systems, the stability of fixed points is often determined by analyzing the eigenvalues of the Jacobian matrix evaluated at the fixed point \cite{bah}\cite{wain}. The eigenvalues provide information about the behavior of perturbations around the fixed point. Consider a system of three autonomous differential equations:
\begin{equation}
\frac{dx}{dN}=f(x,y,z), \,\,\,\frac{dy}{dN}=g(x,y,z),\,\,\, \frac{dz}{dN}=h(x,y,z),\label{eq17}
\end{equation}
with $f$, $g$, and $h$ are functions describing the dynamics of the system. We consider linear perturbations $(\delta x, \,\delta y,\,\delta z)$ as follows 
\begin{equation}
x=x_c+\delta x, \,\,\,\,\, y=y_c+\delta y,\,\,\,\,\, z=z_c+\delta z.\label{eq18}
\end{equation}
Linearizing the autonomous equation above leads to the first-order differential equations
\begin{equation}
{\frac{d}{dN}} 
 \begin{pmatrix}
  \delta x  \\
  \delta y  \\
  \delta z  \\
 \end{pmatrix} = M  \begin{pmatrix}
  \delta x  \\
  \delta y  \\
 \delta z  \\
 \end{pmatrix},\label{eq19}
\end{equation}
where $M$ is $3\times 3$  Jacobian matrix composed of partial derivatives of  $f,\,g,\,h$ with respect to 
$x,\,y,\,z$.  It is given by
\begin{equation}M=
 \begin{pmatrix}
  {\partial f\over \partial x} & {\partial f\over\partial y} & {\partial f\over\partial z}   \\
  {\partial g\over \partial x} & {\partial g \over\partial y} & {\partial g \over\partial z} \\
  {\partial h\over \partial x} & {\partial h \over\partial y} & {\partial h \over\partial z} \\
 \end{pmatrix}_{(x=x_c, \, y=y_c,\,z=z_c)},\label{eq20}
\end{equation}

The stability of the fixed point is determined by the real parts of the eigenvalues. Depending on the signs of the real parts: 1) $Stable$ $node$, if all eigenvalues are real and have negative real parts. 2) $Unstable$ $node$, if all eigenvalues are real and have positive real parts. 3) $Saddle$ $point$, if there is a mix of positive and negative real parts among the eigenvalues. 4) $Stable$ $spiral$, If the real parts of the eigenvalues are negative and at least some of the eigenvalues have non-zero imaginary parts. 5) $Unstable$ $spiral$, if the real parts of the eigenvalues are positive and at least some of the eigenvalues have nonzero imaginary parts \cite{bah}\cite{wain}\cite{col}.

In order to obtain the eigenvalues of each fixed point, we need the explicit form of the potential and the matter coupling function. The choice of potential in dark energy models is crucial because it directly affects the dynamics of the scalar field responsible for the accelerated expansion of the universe. One crucial condition that must be fulfilled is the slow-roll condition \cite{amen}. The slow-roll conditions are more easily satisfied with smaller values of $n$. This choice balances simplicity and the ability to produce a slow-roll regime that can match the observed accelerated expansion of the universe, making it a suitable and popular choice in dark energy models. As a starting step, we choose $n=2$, and for simplification, we also choose the same value for $\alpha$. Then note that there is a special case regarding the fixed points listed in Table  \ref{tabA} for $\alpha=2$. The number of critical points will be reduced to 9 points. Specifically, point P2 coincides with point P3, point P4 coincides with point P5, and points P8 and P9 coincide with point P1. We can classify the stability of a fixed point based on the limitations on the value of the parameter $\omega_{BD}$. The eigenvalues of each fixed point for case $n=\alpha=2$ can be seen in Table \ref{tabB}.
\begin{table}[h]
\caption{Eigenvalue of the fixed points for $n=\alpha=2$.}\label{tabB}%
\begin{tabular}{@{}cccc@{}}
\toprule
Point & $\mu_1$  & $\mu_2$ & $\mu_3$\\
\midrule
P1, P8, P9    & $-\frac{3}{2}$   & $\frac{3}{2}$  & $0$  \\
P2, P3, P4, P5    & $0$   & $\frac{3\left(-6-4\omega_{BD}-G_1\right)}{5\left(3+2\omega_{BD}\right)}$  &  $\frac{3\left(-6-4\omega_{BD}+G_1\right)}{5\left(3+2\omega_{BD}\right)}$\\
P6, P7    & $0$   & $\frac{3\left(H_1-I_1\right)}{5\left(3+2\omega_{BD}\right)\sqrt{-1+2\omega_{BD}}}$  &  $\frac{3\left(H_1+I_1\right)}{5\left(3+2\omega_{BD}\right)\sqrt{-1+2\omega_{BD}}}$  \\
P10, P11    & $-\frac{6 \omega_{BD}-23}{2 (2 \omega_{BD}-1)}$   & $\frac{J_1+K_1}{12(2 \omega_{BD}-1)^3}$  &  $\frac{J_1-M_1}{12(2 \omega_{BD}-1)^3}$  \\
P12    & $ \frac{L_1\left(L_1-3\right)}{4 \omega_{BD}}$   & $ -\frac{3 \left(-3-2 \omega_{BD}+ L_1\right)}{2 \omega_{BD}}$  & $\frac{6 \omega_{BD}-3+L_1}{2 \omega_{BD}}$  \\
P13    & $ \frac{L_1\left(L_1+3\right)}{4 \omega_{BD}}$    & $-\frac{3-6 \omega_{BD}+L_1}{2 \omega_{BD}} $  & $\frac{3 \left(2 \omega_{BD}+3+L_1\right)}{2 \omega_{BD}} $  \\
\botrule
\end{tabular}
\footnotetext{$G_1=\sqrt{24 \omega_{BD}^3-4\omega_{BD}^2-174 \omega_{BD}-171}$}
\footnotetext{$H_1=-6\sqrt{-1+2\omega_{BD}}-4\omega_{BD}\sqrt{-1+2\omega_{BD}}$}
\footnotetext{$I_2=\left(48\omega_{BD}^4-64 \omega_{BD}^3+16 (2 \omega_{BD}-1) \omega_{BD}^2-424 \omega_{BD}^2+48 (2 \omega_{BD}-1) \omega_{BD}-192 \omega_{BD}\right.$}
\footnotetext{$\,\,\,\,\,\,\,\,\,\,\,\,\,\,\left.+36 (2 \omega_{BD}-1)+207 \right)^{1/2}$}
\footnotetext{$J_1=-144 \omega_{BD}^2(2 \omega_{BD}-1) +432\omega_{BD} (2 \omega_{BD}-1) -180 (2 \omega_{BD}-1)$}
\footnotetext{$K_1=\sqrt{147456 \omega_{BD}^4-294912 \omega_{BD}^3+221184 \omega_{BD}^2-73728 \omega_{BD}+9216}$}
\footnotetext{$L_1=\sqrt{3} \sqrt{2 \omega_{BD}+3}$}
\end{table}

Based on the eigenvalues, we can determine the characteristics of each fixed point based on the limits of the parameter $\omega_{BD}$. Point P1, P8 and P9 will always be a saddle point because it does not depend on the parameter $\omega_{BD}$, regardless of any choice of constant values $n$ and $\alpha$. Points P2-P7 will be will be always saddle points for $\omega_{BD}>\frac{23}{6}$. Points P10 and P11 will be stable under condition $\omega_{BD}>\frac{23}{6}$ and will be saddle point for $\omega_{BD}<\frac{23}{6}$. Points P12 and P13 will always be unstable when $\omega_{BD}>0$ and will never be stable.

There is one conditions that permit the existence of stable fixed points. To assess the stability of our mechanism, we introduce $\omega_{BD}^*$ as the threshold value of the Brans-Dicke parameter. We will review one condition, focusing on the limitations of the Brans-Dicke parameter for $\omega_{BD}>\omega_{BD}^*$, with $\omega_{BD}^*=23/6$  for the specific case we have chosen. The stability characteristics of each fixed point for the case $\omega_{BD}>\omega_{BD}^*$ can be seen in Table \ref{tabC} for $\omega_{BD}=4$. Based on the eigenvalue analysis, there are nine saddle points, namely points P1-P9, two stable points, namely points P10 and P11, and two unstable points, namely points P12 and P13. 

\begin{table}[h]
\caption{Stability of each fixed point for $n=\alpha=2$ and $\omega_{BD}=4$}\label{tabC}%
\begin{tabular}{@{}ccccccccccc@{}}
\toprule
Point & $x$ & $y$ & $z$ & $\mu_1$ & $\mu_2$ & $\mu_3$ & $\Omega_{\phi}$ & $\Omega_m$ & $\omega_{eff}$ & Stability\\
\midrule
P1,P8,P9    & $0$ & $0$ & $0$ & $-1.5$ & $1.5$ & $0$ & $0$ & $1$  & 0 & Saddle  \\
P2,P3    & $1.2$ & $-1.149i$ & $0$ & $-2.54$ & $0.142$ & $0$ & $1.079$ & $-0.079$ & $-0.76$ & Saddle  \\
P4,P5    & $1.2$ & $1.149i$ & $0$ & $-2.54$ & $0.142$ & $0$ & $1.079$ & $-0.079$ & $-0.76$ & Saddle \\
P6    & $1.2$ & $-1.149i$ & $0$ & $-2.54$ & $0.142$ & $0$ & $1.079$ & $-0.079$ & $-0.76$ & Saddle  \\
P7    & $1.2$ & $1.149i$ & $0$ & $-2.54$ & $0.142$ & $0$ & $1.079$ & $-0.079$ & $-0.76$ & Saddle  \\
P10    & $1.143$ & $-1.128i$ & $0$ & $-2.43$ & $-0.143$ & $-0.071$ & $1$ & $0$ & $-0.78$& Stable  \\
P11    & $1.143$ & $1.128i$ & $0$ & $-2.43$ & $-0.143$ & $-0.071$ & $1$ & $0$ & $-0.78$ & Stable  \\
P12    & $-0.686$ & $0$ & $0$ & $3.34$ & $1.97$ & $0.98$ & $0.98$ & $0.02$ & $0.54$& Unstable  \\
P13    & $2.186$ & $0$ & $0$ & $6.28$ & $3.14$ & $1.91$ & $0.98$ & $0.02$ & $2.45$& Unstable  \\
\botrule
\end{tabular}
\end{table}

Points P1, P8, and P9  represent the matter-dominated era with $\Omega_m=1$. These points indicate the early universe was dominated by matter. Points P2 to P7 are saddle points but have non-physical density parameters and do not correspond to any physical era. The density parameter values for both the scalar field and matter must not be greater than one and must not be negative. These points highlight non-physical solutions, emphasizing constraints on the model. Points P10 and P11 represent the era of dark energy domination, with $\Omega_{\phi}=1$. The stability of these points suggests that the universe will eventually evolve into a phase dominated by dark energy, leading to accelerated expansion. Points P12 and P13 can be considered as kinetic points of the scalar field with density parameter values approaching 1. The instability indicates that these states are not sustainable and the system will evolve away from these points. 

\begin{figure}[h]
\centering
\includegraphics[width=1\textwidth]{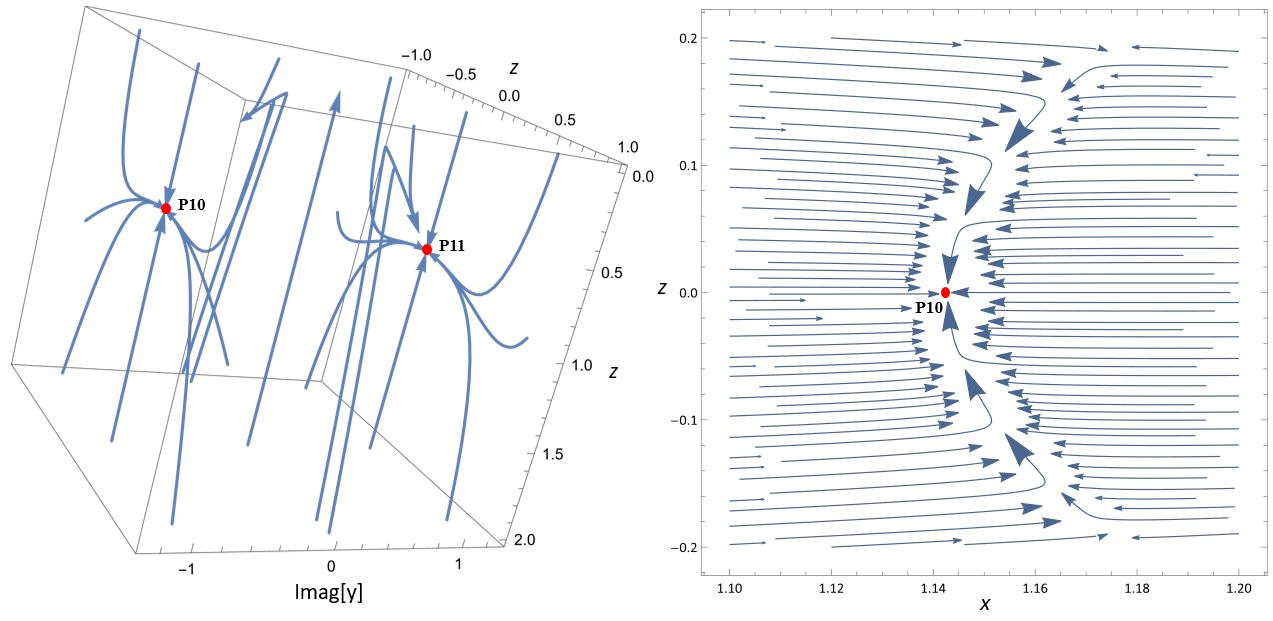}
\caption{(Left) 3D Phase plot of point P10 and P11 for $\omega_{BD}=4$ and $n=\alpha=2$. (Right) Visualization of point P10 in $xz$-plane with set $y=-1.128i$.}\label{figA}
\end{figure}

\begin{figure}[h]
\centering
\includegraphics[width=1\textwidth]{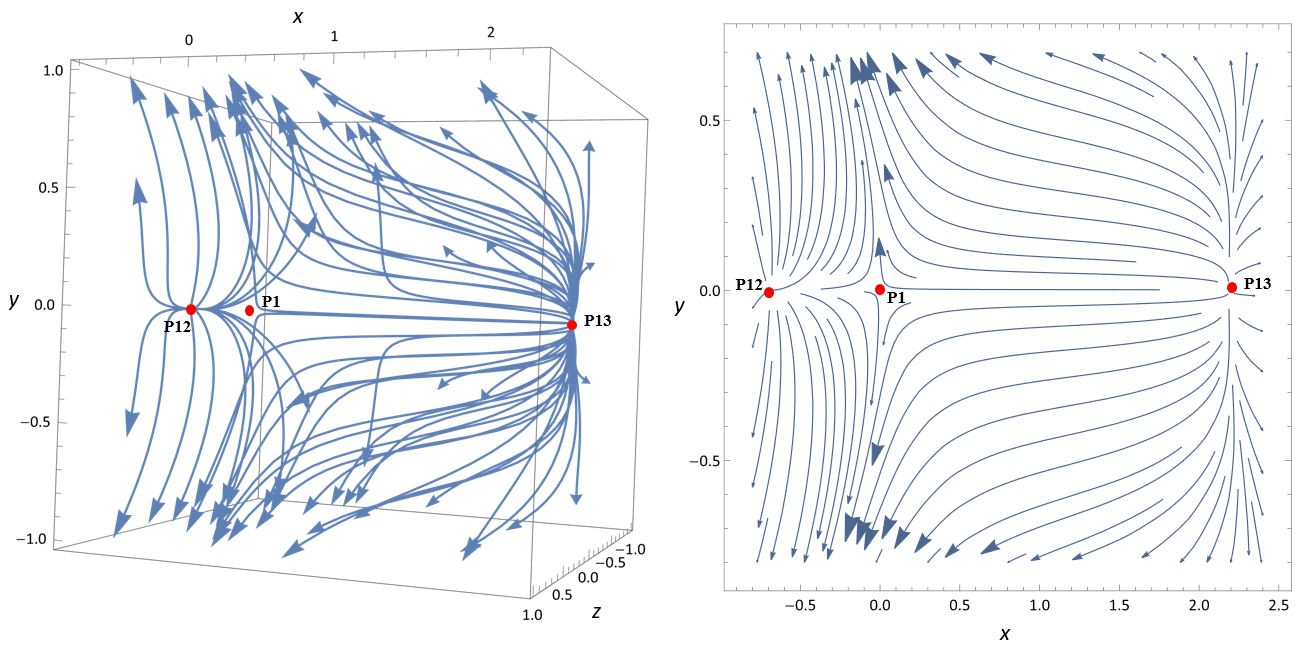}
\caption{(Left) 3D Phase plot of point P1(P8,P9), P12 and P13 for $\omega_{BD}=4$ and $n=\alpha=2$. (Right) Visualization in $xy$-plane with set $z=0$.}\label{figB}
\end{figure}

We can also identify the phase in which a fixed point is situated by analyzing the $\omega_{eff}$ parameter. For the universe to be in an accelerating phase, the effective equation of state parameter $\omega_{eff}$ must be less than $-1/3$. It is evident that points P2-P7 and P10, P11 satisfy the conditions for being in the accelerated phase of the universe. This is further supported by the dominance of the scalar field in these points, as indicated by their density parameter values, particularly for the stable points P10 and P11.

To gain a deeper understanding of the stability characteristics, we can analyze the trajectory of each fixed point in the phase space. Visualizations were performed using Mathematica \cite{Mathematica}. Figure \ref{figA} on the left shows the trajectory paths around points P10 and P11 in 3D phase space. To further confirm the attractor properties of the fixed point, we can set one of the critical point components as constant and visualize it in a 2D plane. Figure \ref{figA} on the right shows the trajectory around point P10 by setting $y=-1.128i$, demonstrating the point's attractor nature in $xz$-plane. A similar plot would be observed for point P11 with $y=1.128i$. In Figure \ref{figB}, we can see the trajectory paths at points P1 (and also P8,P9), P12, and P13. In the $xy$-plane with set $z=0$, the characteristics of each point are clearly visible, with P1 being a saddle point and P12 and P13 being unstable points. Points P2-P7 are not visualized due to their non-physical nature. These points may represent theoretical solutions that do not manifest in real-world scenarios or under specific model constraints, thus not contributing to the stability analysis of the system.

As a comparison, we will examine the case where $n\neq 2$
so that all fixed points exist with different values. Table \ref{tabD} shows the eigenvalues of each critical point for the case $n=3$. Point P1 still be a saddle point. Points P2, P3, P4, and P5 will be stable under the condition $\frac{1}{192}\left(107+27\sqrt{417}\right)\leq\omega_{BD}<\frac{25}{6}$, stable spiral for $0<\omega_{BD}<\frac{1}{192}\left(107+27\sqrt{417}\right)$ and will be saddle points for $\omega_{BD}>\frac{25}{6}$. Points P6 and P7 will always be saddle points under the condition $\omega_{BD}>\frac{19}{6}$. Points P8 and P9 will be always saddle points when $\omega_{BD}<-\frac{53}{24}$ or $\omega_{BD}>-\frac{7}{4}$ and can be stable under certain conditions, specifically when $-\frac{53}{24}<\omega_{BD}<-\frac{7}{4}$. Points P10 and P11 will be stable under condition $\omega_{BD}>\frac{25}{6}$ and will be saddle point for $\omega_{BD}<\frac{25}{6}$. Points P12 and P13 still be unstable when $\omega_{BD}>0$ and will never be stable.

\begin{table}[h]
\caption{Eigenvalue of the fixed points for $n=2$ and $\alpha=3$.}\label{tabD}%
\begin{tabular}{@{}cccc@{}}
\toprule
Point & $\mu_1$  & $\mu_2$ & $\mu_3$\\
\midrule
P1    & $-\frac{3}{2}$   & $\frac{3}{2}$  & $0$  \\
P2, P3, P4, P5    & $-\frac{3}{11}$   & $\frac{-81-54\omega_{BD}-G_2}{22\left(3+2\omega_{BD}\right)}$  &  $\frac{-81-54\omega_{BD}+G_2}{22\left(3+2\omega_{BD}\right)}$\\
P6, P7    & $\frac{3}{20}$   & $\frac{3\left(H_2-I_2\right)}{5\left(3+2\omega_{BD}\right)\sqrt{-1+2\omega_{BD}}}$  &  $\frac{3\left(H_2+I_2\right)}{5\left(3+2\omega_{BD}\right)\sqrt{-1+2\omega_{BD}}}$  \\
P8, P9    & $ \frac{24\omega_{BD}+53}{4 (4\omega_{BD}+7)} $   & $\frac{J_2-K_2}{480(4\omega_{BD}+7)^3}$ &  $\frac{J_2+K_2}{480(4\omega_{BD}+7)^3}$  \\
P10, P11    & $-\frac{6 \omega_{BD}-25}{2 (2 \omega_{BD}-1)}$   & $\frac{L_2+M}{12(2 \omega_{BD}-1)^3}$  &  $\frac{L_2-M}{12(2 \omega_{BD}-1)^3}$  \\
P12    & $ -\frac{3\left(-3-2\omega_{BD}+N\right)}{2 \omega_{BD}}$   & $ \frac{-3+6\omega_{BD}+N}{2 \omega_{BD}}$  & $\frac{\left(N-3\right)\left(N-1\right)}{8 \omega_{BD}}$  \\
P13    & $ -\frac{3-6\omega_{BD}+N}{2 \omega_{BD}}$    & $ \frac{3\left(3+2\omega_{BD}+N\right)}{2 \omega_{BD}}$  & $\frac{\left(N+3\right)\left(N+1\right)}{8 \omega_{BD}}$  \\
\botrule
\end{tabular}
\footnotetext{$G_2=\sqrt{3}\sqrt{1152\omega_{BD}^3+444 \omega_{BD}^2-11068\omega_{BD}-13713}$}
\footnotetext{$H_2=-6\sqrt{-1+2\omega_{BD}}-4\omega_{BD}\sqrt{-1+2\omega_{BD}}$}
\footnotetext{$I_2=\left(48\omega_{BD}^4-64 \omega_{BD}^3+16 (2 \omega_{BD}-1) \omega_{BD}^2-424 \omega_{BD}^2+48 (2 \omega_{BD}-1) \omega_{BD}-192 \omega_{BD}\right.$}
\footnotetext{$\,\,\,\,\,\,\,\,\,\,\,\,\,\,\left.+36 (2 \omega_{BD}-1)+207 \right)^{1/2}$}
\footnotetext{$J_2= \left(-5760\omega_{BD}^2-18000\omega_{BD}-13860\right)(4\omega_{BD}+7)$}
\footnotetext{$K_2=\left(530841600\omega_{BD}^6+5352652800 \omega_{BD}^5+22473216000\omega_{BD}^4+50287104000  \omega_{BD}^3\right.$}
\footnotetext{$\,\,\,\,\,\,\,\,\,\,\,\,\,\,\left.+63249984000 \omega_{BD}^2+42398092800  \omega_{BD}+11833088400\right)^{1/2}$}
\footnotetext{$L_2=-144 \omega_{BD}^2(2 \omega_{BD}-1) +432\omega_{BD} (2 \omega_{BD}-1) -180 (2 \omega_{BD}-1)$}
\footnotetext{$M=\sqrt{147456 \omega_{BD}^4-294912 \omega_{BD}^3+221184 \omega_{BD}^2-73728 \omega_{BD}+9216}$}
\footnotetext{$N=\sqrt{3} \sqrt{2 \omega_{BD}+3}$}
\end{table}

There are three conditions that permit the existence of stable fixed points, but not all of them will be examined. In the Brans-Dicke theory, the scalar field $\phi$ contributes to gravitational interactions that the kinetic term of the scalar field in action has the correct sign, corresponding to positive kinetic energy. Negative kinetic energy would lead to instabilities, such as ghost fields, which are unphysical. A crucial step in achieving this is by ensuring that the parameter $\omega_{BD}$ is positive. We will review only two conditions, focusing on the limitations of the Brans-Dicke parameter for $\omega_{BD}>\omega_{BD}^*$ and $0<\omega_{BD}<\omega_{BD}^*$, with $\omega_{BD}^*=25/6$  for the specific case we have chosen.

\subsection{Case $\omega_{BD}>\omega_{BD}^*$}

The stability characteristics of each fixed point for the case $\omega_{BD}>\omega_{BD}^*$ can be seen in Table \ref{tabE} for $\omega_{BD}=5$. Based on the eigenvalue analysis, there are nine saddle points, namely points P1-P9, two stable points, namely points P10 and P11, and two unstable points, namely points P12 and P13. 

\begin{table}[h]
\caption{Stability of each fixed point for $n=2$, $\alpha=3$ and $\omega_{BD}=5$}\label{tabE}%
\begin{tabular}{@{}ccccccccccc@{}}
\toprule
Point & $x$ & $y$ & $z$ & $\mu_1$ & $\mu_2$ & $\mu_3$ & $\Omega_{\phi}$ & $\Omega_m$ & $\omega_{eff}$ & Stability\\
\midrule
P1    & $0$ & $0$ & $0$ & $-1.5$ & $1.5$ & $0$ & $0$ & $1$  & 0 & Saddle  \\
P2    & $1.09$ & $-1.196i$ & $-0.182$ & $-3.004$ & $0.549$ & $-0.273$ & $1.33$ & $-0.33$ & $-0.79$ & Saddle  \\
P3    & $1.09$ & $-1.196i$ & $0.182$ & $-3.004$ & $0.549$ & $-0.273$ & $1.33$ & $-0.33$ & $-0.79$ & Saddle  \\
P4    & $1.09$ & $1.196i$ & $-0.182$ & $-3.004$ & $0.549$ & $-0.273$ & $1.33$ & $-0.33$ & $-0.79$ & Saddle \\
P5    & $1.09$ & $1.196i$ & $0.182$ & $-3.004$ & $0.549$ & $-0.273$ & $1.33$ & $-0.33$ & $-0.79$ & Saddle  \\
P6    & $1.2$ & $-1.25i$ & $0$ & $-3.19$ & $0.79$ & $0.15$ & $1.56$ & $-0.56$ & $-0.76$ & Saddle  \\
P7    & $1.2$ & $1.25i$ & $0$ & $-3.19$ & $0.79$ & $0.15$ & $1.56$ & $-0.56$ & $-0.76$ & Stable  \\
P8    & $-0.074$ & $0$ & $-0.303i$ & $1.602$ & $-1.435$ & $0.019$ & $0.078$ & $0.922$ & $-0.006$ & Saddle  \\
P9    & $-0.074$ & $0$ & $0.303i$ & $1.602$ & $-1.435$ & $0.019$ & $0.078$ & $0.922$ & $-0.006$ & Saddle  \\
P10    & $0.89$ & $-1.109i$ & $0$ & $-2.56$ & $-0.78$ & $-0.28$ & $1$ & $0$ & $-0.87$& Stable  \\
P11    & $0.89$ & $1.109i$ & $0$ & $-2.56$ & $-0.78$ & $-0.28$ & $1$ & $0$ & $-0.87$& Stable  \\
P12    & $-0.65$ & $0$ & $0$ & $3.32$ & $2.03$ & $0.93$ & $0.98$ & $0.02$ & $0.57$& Unstable  \\
P13    & $1.85$ & $0$ & $0$ & $5.77$ & $3.12$ & $2.076$ & $0.98$ & $0.02$ & $2.23$& Unstable  \\
\botrule
\end{tabular}
\end{table}

Point P1 represents the matter-dominated era with $\Omega_m=1$. These points indicate the early universe was dominated by matter. Points P2-P7 have non-physical density parameters and do not correspond to any physical era. Points P8 and P9 represent an era where the scalar field starts moving but matter still dominates, with $\Omega_{m}=0.922$. These points suggest transitional eras where both matter and scalar fields are significant. Points P10 and P11 represent the era of dark energy domination, with $\Omega_{\phi}=1$. The stability of these points implies that the universe will eventually transition into a dark energy-dominated phase, resulting in accelerated expansion. Points P12 and P13 can be considered as kinetic points of the scalar field with density parameter values approaching 1. The instability shows that these states are not sustainable, and the system will move away from these points over time. 

\begin{figure}[h]
\centering
\includegraphics[width=1\textwidth]{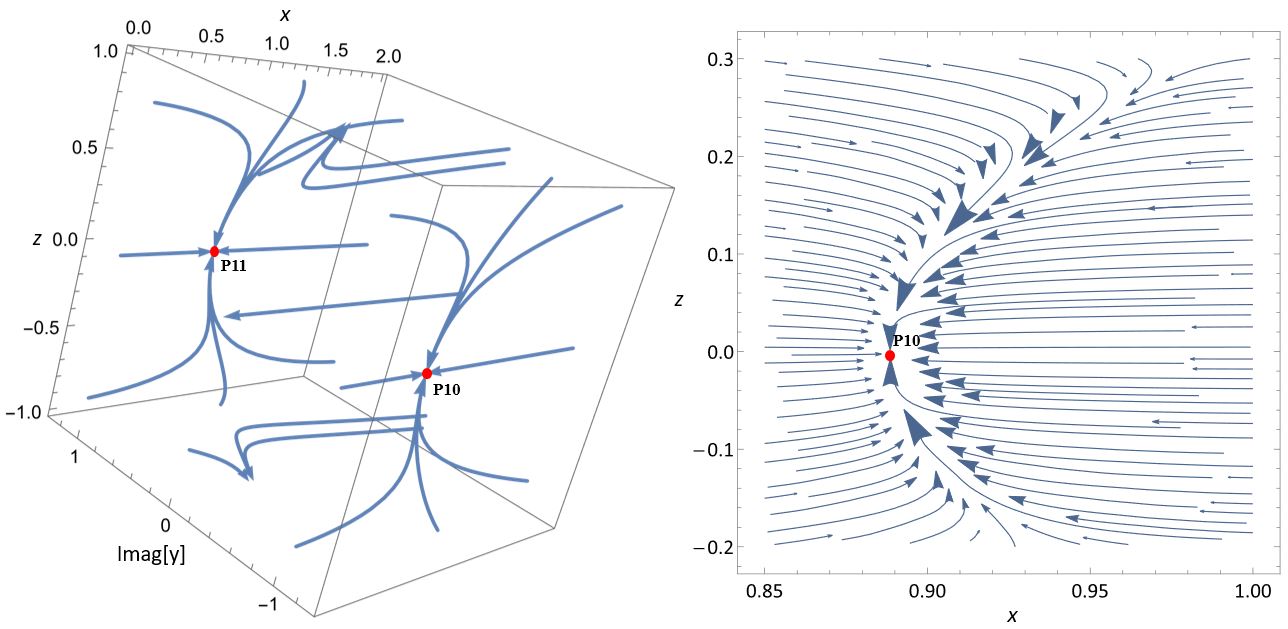}
\caption{(Left) 3D Phase plot of point P10 and P11 for $n=2$, $\alpha=3$ and $\omega_{BD}=5$. (Right) Visualization of point P10 in $xz$-plane with set $y=-1.109i$.}\label{figC}
\end{figure}

We can also determine the phase in which a fixed point resides by examining the $\omega_{eff}$ parameter. For the universe to be in an accelerating phase, the effective equation of state parameter $\omega_{eff}$ must be less than $-1/3$. It is clear that points P2-P7 and P10, P11 meet the criteria for being in the accelerated phase of the universe. This conclusion is reinforced by the dominance of the scalar field at these points, as reflected in their density parameter values, especially for the stable points P10 and P11.

\begin{figure}[h]
\centering
\includegraphics[width=1\textwidth]{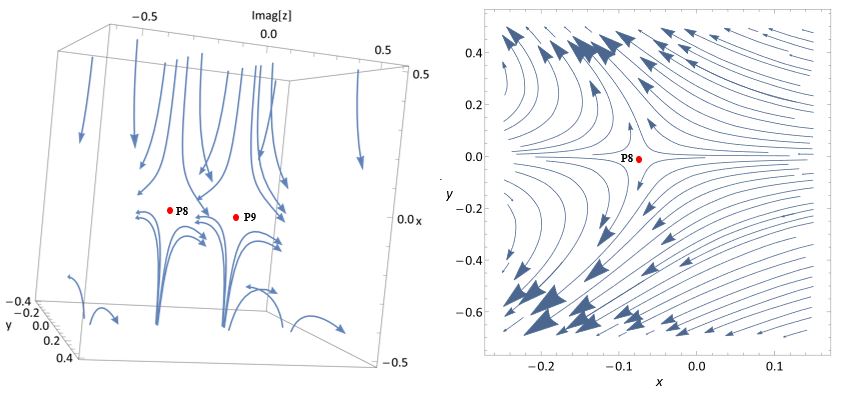}
\caption{(Left) 3D Phase plot of point P8 and P9 for $n=2$, $\alpha=3$ and $\omega_{BD}=5$. (Right) Visualization in $xy$-plane of point P8 with set $z=-0.303i$.}\label{figD}
\end{figure}

\begin{figure}[h]
\centering
\includegraphics[width=1\textwidth]{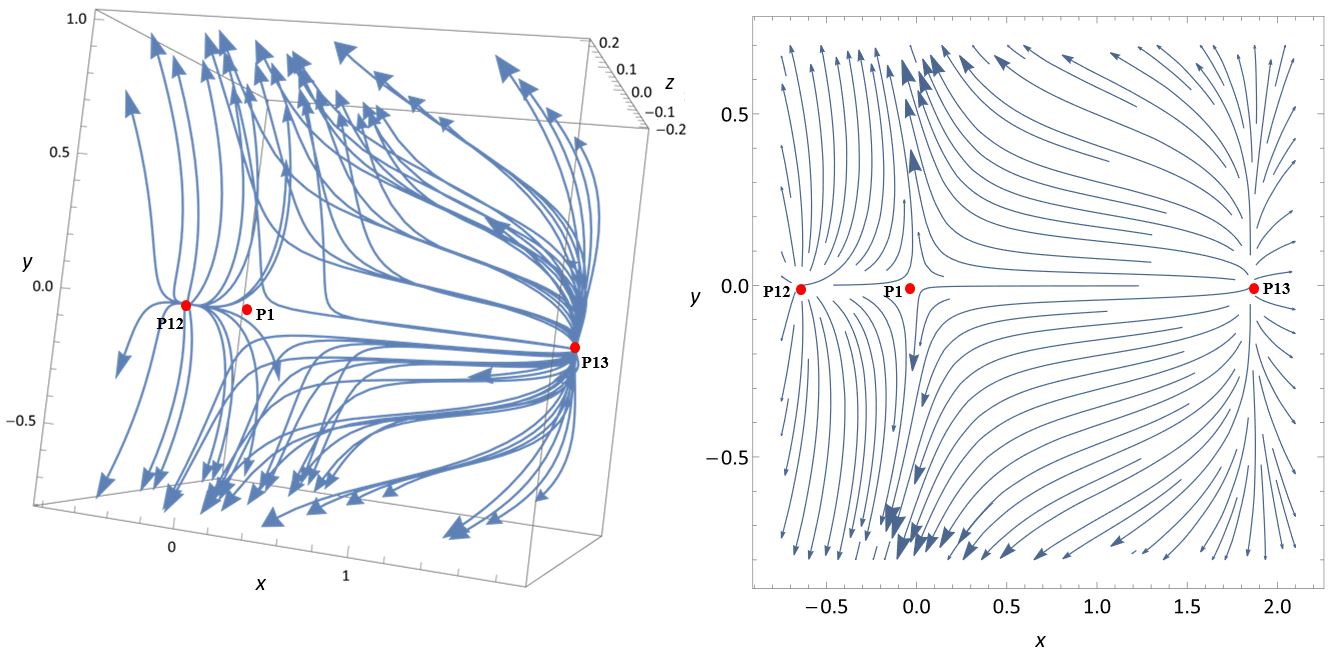}
\caption{(Left) 3D Phase plot of point P1, P12 and P13 for $n=2$, $\alpha=3$ and $\omega_{BD}=5$. (Right) Visualization in $xy$-plane with set $z=0$.}\label{figE}
\end{figure}

To better comprehend the stability properties, we can analyze the behavior of each fixed point in the phase space. Figure \ref{figC} on the left shows the trajectory paths around points P10 and P11 in 3D phase space. To further confirm the attractor properties of the fixed point, we can set one of the critical point components as constant and visualize it in a 2D plane. Figure \ref{figC} on the right shows the trajectory around point P10 by setting $y=-1.109i$, demonstrating the point's attractor nature in $xz$-plane. A similar plot would be observed for point P11 with $y=1.109i$. Figure \ref{figD} shows the phase space plot of points P8 and P9, with figures on the left providing 3D visual representations. The saddle nature of point P8 is further confirmed in figures on the right when the system is observed from an alternative viewpoint in the $xy$-plane with set $z= -0.303i$. Figure \ref{figE} shows the trajectory paths at points P1, P12, and P13. In the $xy$-plane with set $z=0$, the characteristics of each point are clearly visible, with P1 being a saddle point and P12 and P13 being unstable points. Points P2-P7 are not visualized due to their non-physical nature. These points might be theoretical solutions that fail to appear in real-world scenarios or under certain model constraints, thereby not contributing to the stability analysis of the system.

\subsection{Case $0<\omega_{BD}<\omega_{BD}^*$}

We have chosen the value $\omega_{BD}=4$ which falls within the range $0<\omega_{BD}<\omega_{BD}^*$, to examine the characteristics of each fixed point in the context of the stability of the dark energy model. According to the eigenvalues listed in Table \ref{tabF}, there are 4 stable points, 2 unstable points, and the remainder are saddle points. Interestingly, some fixed points that were initially stable have become saddle points, and vice versa. Points P1, P6-P9, P12, and P13 retain the same characteristics as they did with the previous selection of $\omega_{BD}$.  Points P2-P5, previously identified as saddle points, are now stable with all negative eigenvalues and remain in the accelerated phase based on their $\omega_{eff}$ value. However, their density parameter values do not align with the expected properties of stable points, specifically the absolute dominance of the scalar field. Although the scalar field remains dominant with $\Omega_{\phi}=0.93$, there is still a non-zero matter density, typically a characteristic of saddle points. Conversely, points P10 and P11, previously stable, have transitioned to saddle points based on their eigenvalues, yet they still remain in the accelerated phase. Despite this, their density parameter values indicate the total dominance of the scalar field with $\Omega_{\phi}=1$ which is usually a feature of stable fixed points.  

\begin{table}[h]
\caption{Stability of each fixed point for $n=2$, $\alpha=3$ and $\omega_{BD}=4$}\label{tabF}%
\begin{tabular}{@{}ccccccccccc@{}}
\toprule
Point & $x$ & $y$ & $z$ & $\mu_1$ & $\mu_2$ & $\mu_3$ & $\Omega_{\phi}$ & $\Omega_m$ & $\omega_{eff}$ & Stability\\
\midrule
P1    & $0$ & $0$ & $0$ & $-1.5$ & $1.5$ & $0$ & $0$ & $1$ & $0$ & Saddle  \\
P2    & $1.09$ & $-1.109i$ & $-0.08i$ & $-2.31$ & $-0.27$ & $-0.145$ & $0.93$ & $0.07$  & $-0.804$ & Stable  \\
P3    & $1.09$ & $-1.109i$ & $0.08i$ & $-2.31$ & $-0.27$ & $-0.145$ & $0.93$ & $0.07$  & $-0.804$ & Stable  \\
P4    & $1.09$ & $1.109i$ & $-0.08i$ & $-2.31$ & $-0.27$ & $-0.145$ & $0.93$ & $0.07$  & $-0.804$ & Stable \\
P5    & $1.09$ & $1.109i$ & $0.08i$ & $-2.31$ & $-0.27$ & $-0.145$ & $0.93$ & $0.07$  & $-0.804$ & Stable  \\
P6    & $1.2$ & $-1.149i$ & $0$ & $-2.54$ & $0.15$ & $0.142
$ & $1.08$ & $-0.08$ & $-0.76$ & Saddle  \\
P7    & $1.2$ & $1.149i$ & $0$ & $-2.54$ & $0.15$ & $0.142$ & $1.08$ & $-0.08$ & $-0.76$ & Saddle  \\
P8    & $-0.087$ & $0$ & $-0.301i$ & $1.619$ & $-1.424$ & $0.022$ & $0.092$ & $0.908$ & $-0.007$ & Saddle  \\
P9    & $-0.087$ & $0$ & $0.301i$ & $1.619$ & $-1.424$ & $0.022$ & $0.092$ & $0.908$ & $-0.007$ & Saddle  \\
P10    & $1.143$ & $-1.128i$ & $0$ & $-2.43$ & $-0.143$ & $0.071$ & $1$ & $0$ & $-0.78$ & Saddle \\
P11    & $1.143$ & $1.128i$ & $0$ & $-2.43$ & $-0.143$ & $0.071$ & $1$ & $0$ & $-0.78$ & Saddle  \\
P12    & $-0.686$ & $0$ & $0$ & $3.343$ & $1.97$ & $0.89$ & $1$ & $0$ & $0.54$ & Unstable  \\
P13    & $2.186$ & $0$ & $0$ & $6.28$ & $3.413$ & $1.907$ & $1$ & $0$ & $2.45
$ & Unstable  \\
\botrule
\end{tabular}
\end{table}

\subsection{Variation of $\omega_{BD}$}

To gain a clearer understanding of the stability changes in two groups of fixed points that can potentially be stable, namely the groups of points (P2-P5) and (P10, P11), we can examine various conditions with different values of $\omega_{BD}$ values within the two ranges mentioned above. Table \ref{tabG} shows the stability of the fixed points for $\omega_{BD}=2$ and $\omega_{BD}=3$, which fall within the range $0<\omega_{BD}<\omega_{BD}^*$, and for $\omega_{BD}=10$, $\omega_{BD}=10^3$ and $\omega_{BD}=10^5$ which fall within the range $\omega_{BD}>\omega_{BD}^*$. In the range $0<\omega_{BD}<\omega_{BD}^*$, points P2-P5 consistently show stability, as stable spiral points, while points P10 and P11 consistently exhibit saddle behavior. However, this stability is not supported by their density parameter values. In fact, as $\omega_{BD}$ decreases, the matter density becomes more dominant, leading to a contradiction between stability based on eigenvalues and stability based on the density parameters of dark energy and matter. In the range $\omega_{BD}>\omega_{BD}^*$ points P2-P5 consistently exhibit saddle behavior, and points P10 and P11 consistently show stability. This is corroborated by their density parameter values, where the scalar field dominates stable conditions. This analysis sheds light on how the stability properties of fixed points in the dark energy model change with different values of $\omega_{BD}$, providing insight into the complex dynamics and stability conditions influenced by the Brans-Dicke parameter.
\begin{table}[h]
\caption{Comparison of the stability of fixed points across several variations of $\omega_{BD}$.}\label{tabG}%
\begin{tabular}{@{}cccccccc@{}}
\toprule
$\omega_{BD}$ & Point & $\mu_1$ & $\mu_2$ & $\mu_3$ & $\Omega_{\phi}$ & $\Omega_m$  &Stability\\
\midrule
  2 &  P2,P3,P4,P5  & $-1.23+1.77i$ & $-1.23-1.77i$ & -0.27 & 0.14 & 0.86 & Stable Spiral \\
    &  P10,P11 & 3.67 & 2.17 & -1.67 &  1 & 0 & Saddle  \\
\midrule
 3  &  P2,P3,P4,P5  & $-1.23+0.95i$ & $-1.23-0.95i$ & -0.27 & 0.718 & 0.282 & Stable Spiral \\
    &  P10,P11 & -2.54 & 0,9 & 0,14 & 1 & 0 & Saddle  \\
\midrule
  $10$ &  P2,P3,P4,P5  & -4.77 & 2.32 & -0.27 & 3.314 & -2.314 & 
  Saddle  \\
   &  P10,P11 & -2.79 & -1.95 & -0.92 & 1 & 0 & Stable  \\  
\midrule
$10^3$ &  P2,P3,P4,P5  & -43.42 & 40.97 & -0.27 & 396,04 & -395.04 & Saddle  \\
   &  P10,P11 & -2.99 & -2.98 & -1.49 & 1 & 0 & Stable  \\
\midrule
$10^5$  &  P2,P3,P4,P5  & -423.73 & 421.27 & -0.27 & 39668.71 & -39667.71 & Saddle  \\
   &  P10,P11 & -2.99 & -2.99 & -1.49 & 1 & 0 & Stable  \\   
\botrule
\end{tabular}
\end{table}
\subsection{Variation of $n$ and $\alpha$}
In the previous work, we reviewed the stability of fixed points for specific values of $n$ and $\alpha$, which led to certain constraints on the Brans-Dicke parameter. In this section, we aim to vary the values of  $n$ and $\alpha$, exploring different forms of the potential and the matter coupling function. Table \ref{tabH} presents several combinations, specifically for  $n=2$, $n=3$, $n=4$ and $\alpha=4$, $\alpha=5$, $\alpha=6$. Each combination of $n$ and $\alpha$ results in a unique value of $\omega_{BD}^*$. As the values of both of them increase, the value of $\omega_{BD}^*$ also increases.  This implies that each choice of the potential and matter coupling function imposes different constraints on the Brans-Dicke parameter $\omega_{BD}$. These constraints are crucial in ensuring the stability of each fixed point in the dark energy model under consideration.

\begin{table}[h]
\caption{Several values of $\omega_{BD}^*$ based on variations in $n$ and $\alpha$}\label{tabH}%
\begin{tabular}{@{}cccccccccc@{}}
\toprule
 $n$ & 2 & 2 & 2 & 3 & 3 & 3 & 4 & 4 & 4  \\
\midrule
  $\alpha$ & 4 & 5 & 6 & 4 & 5 & 6 & 4 & 5 & 6   \\
\midrule
$\omega_{BD}^*$ & $\frac{9}{2}$ & $\frac{29}{6}$ & $\frac{31}{6}$ & $\frac{23}{3}$ & $\frac{97}{12}$ & $\frac{17}{2}$ & $\frac{23}{2}$ & $12$ & $\frac{25}{2}$\\

\botrule
\end{tabular}
\end{table}

\section{Brans-Dicke Chameleon and the Acceleration of The Universe's Expansion}\label{sec5}
In the case examined above, we identified several stable fixed points for specific values of $\omega_{BD}$. For the special case when $\alpha=2$, stability is achieved when $\omega_{BD}>\omega_{BD}^*$. For the case $\alpha \neq 2$, and as an example $\alpha=3$ as mentioned above, the stable fixed points are divided into two groups: one where $\omega_{BD}>\omega_{BD}^*$ and another where $0<\omega_{BD}<\omega_{BD}^*$, with $\omega_{BD}^*=25/6$. Both groups of fixed points are in the accelerated expansion phase of the universe with $\omega_{eff}<-\frac{1}{3}$. In the first group, we found fixed points that are stable based on their eigenvalues and are supported by density parameter values for dark energy and matter that match the actual conditions of the universe. In the second group, we also found stable fixed points based on their eigenvalues, which contradict the density parameter values. Therefore, we will focus on the first group, which can be considered as physically stable points.
It is established that the variable $x$ denotes the kinetic term of the scalar field, $y$ represents the potential term of the scalar field, and $z$ signifies the coupling term between the matter density and the scalar field. The obtained results lead to the inference that the stability of the Chameleon mechanism in the Brans-Dicke model manifests when the scalar field is dominant especially the potential terms, accompanied by the minimum coupling of matter and scalar field.

In the discourse surrounding modified gravity, specific models posit the scalar field as a potential contender for dark energy, as seen in Quintessence, Gauss-Bonnet, and similar frameworks \cite{bah}. Essentially, the attractor solutions derived earlier suggest that stability is achieved when the scalar field takes on a dominant role, signifying the prevalence of dark energy. Upon examining other dark energy models that also put forth the scalar field as a candidate, it becomes apparent that the attractor solutions in those models align with the dominance of the scalar field, consistent with the aforementioned results.

The dominance of the scalar field’s kinetic and potential terms in the universe’s acceleration is a concept rooted in the dynamics of specific cosmological models, particularly those involving scalar fields. The kinetic term represents the rate of change of the scalar field for a time. It is associated with the field’s motion or velocity. When the kinetic term dominates, it implies that the scalar field is dynamically evolving, and its motion contributes to the overall energy density of the universe. The potential term represents the energy associated with the scalar field at a given point in space. It characterizes the field’s potential to influence cosmic evolution. When the potential term dominates, it suggests that the scalar field has settled into a particular energy state, potentially leading to the acceleration of the universe's expansion. 

In the specific context of the Chameleon mechanism within the Brans-Dicke model and similar scenarios, the dominance of the scalar field signifies stability. It indicates that the scalar field’s dynamics are primarily influenced by its motion and energy state, with minimal interaction with matter. This dominance contributes to the stability of the Chameleon mechanism in regions of low matter density. Fundamentally, the role of the scalar field in propelling the universe’s acceleration stems from their collaborative capacity to produce an effective energy density characterized by negative pressure, resembling the attributes of dark energy. When appropriately balanced, these terms can lead to the observed accelerated expansion, as seen in certain cosmological models.

A prior investigation identified thin-shell solutions for the Chameleon Brans-Dicke field \cite{sut}. The findings indicated that the scalar field maintains a negligible value in proximity to regions with high matter density (near the matter source). It then experiences an exponential increase, ultimately converging to a stable value in areas of low matter density situated far from the matter source $\left(r\rightarrow\infty\right)$. In locales with high matter density, the Chameleon mechanism conceals the gravitational modification effects. Conversely, in regions with low density, the scalar field reassumes its role as dark energy, instigating the acceleration of the universe's expansion. Crucially, the scalar field reverts to functioning as dark energy when it attains a very large, stable, and constant value, coinciding with very low matter density. Implicitly, these conditions lend support to the attributes of the attractor solutions outlined in the dynamic system analysis discussed earlier.

\section{Conclusion}\label{sec6}

The results of this study provide crucial support for the stability of the Chameleon mechanism, especially within the Brans-Dicke model. By examining the eigenvalues obtained, we define a constraint on the Brans-Dicke parameter, referred to as $\omega_{BD}^*$. This constraint divides the stability points into two distinct groups: one for $\omega_{BD}>\omega_{BD}^*$ and another $0<\omega_{BD}<\omega_{BD}^*$, except for the special case of $\alpha=2$ which only satisfies the conditions of the first group. In the first group, the fixed point is stable and is supported by suitable eigenvalues and a density parameter that represents the era of dark energy dominance. In the second group, the fixed point is stable according to its eigenvalues, but the density parameter does not correspond to the expected conditions of a stable universe. We then explore variations in the explicit forms of the scalar field potential and the matter coupling function by selecting different values of $n$ and $\alpha$. It is shown that each unique constant value results in a distinct $\omega_{BD}^*$. This demonstrates that the stability of the dark energy model is sensitive to the specific values of these parameters. The exploration reveals that the stability conditions for dark energy dominance are closely linked to the chosen values of $n$ and $\alpha$. These parameters influence the behavior of the scalar field potential and the matter coupling function, thereby affecting the overall dynamics of the universe. By identifying the unique $\omega_{BD}^*$ for different parameter values, we gain insights into the stability characteristics of various dark energy models, which is crucial for understanding the long-term evolution of the universe under the influence of dark energy.

From the dynamic analysis performed, it was also discovered that stability is achieved when the scalar field assumes a dominant role, with a particular emphasis on the significance of the kinetic and potential terms. This dominance is highlighted while simultaneously minimizing the influence of matter density in the cosmic environment. In regions of high matter density, the scalar field’s negligible presence effectively conceals the gravitational modification effects, aligning with standard gravitational behavior. However, in areas of low matter density, far from massive objects, the scalar field experiences exponential growth. This growth contributes to the manifestation of dark energy, ultimately leading to the acceleration of the universe. These findings not only affirm the stability of the Chameleon mechanism in the Brans-Dicke model but also provide a deeper understanding of how this mechanism operates in different cosmic environments. The interplay between the scalar field, matter density, and the terms in the gravitational equations contributes valuable insights into the intricate dynamics of the universe on both large and small scales.

\section*{Declarations}

\begin{itemize}
\item Funding\\
This work was supported by Riset Unggulan ITB 2023
\item Competing interests \\
The authors have no relevant financial or non-financial interests to disclose.
\item Author contribution\\
All authors contributed to the study conception and design. Material preparation, data collection and analysis were performed by Azwar Sutiono, Agus Suroso and Freddy Permana Zen. The first draft of the manuscript was written by Azwar Sutiono and all authors commented on previous versions of the manuscript. All authors read and approved the final manuscript
\end{itemize}

\noindent

\nocite{*}
\bibliography{sn-bibliography}

\end{document}